\documentclass[aps,prd,preprint,showpacs,showkeys,preprintnumbers,amsmath,amssymb]{revtex4-1}
\usepackage{amsmath}
\usepackage{amssymb}

\begin{document}

\title{Noncommutative Quantum Mechanics based in Representations of Exotic Galilei Group}

\author{R. G. G. Amorim }
\email{ronniamorim@gmail.com} \affiliation{Instituto de F\'{i}sica,
Universidade de Bras\'{i}lia, 70910-900, Bras\'{i}lia, DF, Brazil.\\
Faculdade Gama, Universidade de Bras\'{i}lia, 72444-240, Setor Leste
(Gama), Bras\'{i}lia, DF, Brazil.}

\author{S. C. Ulhoa }
\email{sc.ulhoa@gmail.com} \affiliation{Instituto de F\'{i}sica,
Universidade de Bras\'{i}lia, 70910-900, Bras\'{i}lia, DF, Brazil.\\
Faculdade Gama, Universidade de Bras\'{i}lia, 72444-240, Setor Leste
(Gama), Bras\'{i}lia, DF, Brazil.}

\begin{abstract}
Using elements of symmetry, we constructed the Noncommutative
Schr\"odinger Equation from a representation of Exotic Galilei
Group. As consequence, we derive the Ehrenfest theorem using
noncommutative coordinates. We also have showed others features of
quantum mechanics in such a manifold. As an important result, we
find out that a linear potential in the noncommutative Schr\"odinger
equation is completely analogous to the ordinary case. We also
worked with harmonic and anharmonic oscillators, giving corrections
in the energy for each one.
\end{abstract}


\maketitle

\section{Introduction}

The concept of continuous group was introduced by the Norwegian
mathematician Sofus Lie~\cite{olver} while working with differential
equations in second half of nineteenth century. Then it has gained
great visibility in 1928 with the work of H. Weyl~\cite{weyl}, who
linked group theory to atomic spectroscopy. Thus it was established
for the first time a direct relation between quantum theory and
group theory. However its importance increased enormously when
Wigner applied the theory of representations of Lie groups to
physical systems. Particularly fundamental particles can be
described by means of irreducible representations of kinematical
groups, such as De Sitter, Galilei and Poincar\'e groups. Hence
fundamental interactions are implemented by gauge symmetries. Later,
in order to find an appropriate Hamiltonian structure for atomic
systems, P. Dirac also used the theory of representation of Lie
groups to investigate such systems~\cite{dirac}. Therefore group
theory, which is an important branch of mathematics, became an
indispensable tool to understand and formalize symmetries in
physics.

The notion of non-commutativity began in physics with the uncertain
principle which is a well known relation between coordinates and
momenta operators. In a letter addressed to Peierls in $1930$,
Heisenberg considered that this very feature could be valid also
between coordinates operators, in such a way that it could precludes
the well known divergences of the self-energy terms calculated for
particles. Since then the idea of noncommutative spatial coordinates
propagated throughout the scientific community until it reached H.
Snyder, a former Ph.D. student of Oppenheimer~\cite{pauli1, pauli2,
jackiw}. Thus, at least formally, Snyder was the first one who
stated that spatial coordinates would not commutate with each other
at small distances, in his work published in $1947$~\cite{snyder1,
snyder2}. He proposed a whole new paradigm in which the spacetime
should be understood as a collection of tiny cells of minimum size,
resulting into a misleading concept of point. Thus, once the minimum
size is reached, in the realm of some high energy phenomenon, the
position should be given by the noncommutative coordinate operators.
As a direct consequence it would be impossible to precisely measure
a particle position.

After Snyder's pioneering, those ideas remained forgotten, mainly
because the success achieved by the so called renormalization
procedure, which has led to an entire new branch of research in
physics, on dealing with ultra-violet divergences of scattering
integrals in the context of quantum field theory. Then in the last
years the interest of the scientific community on noncommutative
geometry has increased due to works on non-abelian
theories~\cite{chans}, gravitation~\cite{kalau, kastler, connes1},
standard model~\cite{connes2, varilly1, varilly2} and on quantum
Hall effect~\cite{belissard}. More recently the discovery that the
dynamics of an open string can be associated with noncommutative
spaces has contributed to the last revival of noncommutative
theories~\cite{witten}.

From the mathematical point of view, the simplest algebra obeyed by
the coordinates operators $\widehat{x}^{\mu}$ is
$$
[\widehat{x}^{\mu},\widehat{ x}^{\nu}]=i\theta^{\mu\nu}\,,
$$
where $\theta^{\mu\nu}$ is a skew-symmetric constant tensor. It
worths to point out that the mean values of the position operators
do correspond to the actual position observed, thus it is said that
such operators are hermitian ones. It is well known in quantum
mechanics that a noncommutative relation between two operators lead
to a specific uncertain relation, hence the above expression yields
$$
\Delta \widehat{x}^{\mu}\Delta \widehat{x}^{\nu}\geq
\frac{1}{2}|\theta^{\mu\nu}|\,.
$$
Following the ideas introduced by Snyder, it possible to associate
the minimum size with a distance of the $\sqrt{|\theta^{\mu\nu}|}$
order of magnitude. Thus the noncommutative effects turn out to be
relevant at such scales. Usually the non-commutativity is introduced
by means of the Moyal product defined as~\cite{Akofor:2008ae}

\begin{small}
\begin{eqnarray}
f(x) \star g(x) &\equiv& \exp
\left(  {i \over 2} \theta^{\mu\nu}  {\partial \over \partial x^\mu}
{\partial \over \partial y^\nu} \right) f(x) g(y) |_{y \rightarrow x} \nonumber\\
&=& f(x) g(x) + {i\over 2} \theta^{\mu \nu}
\partial_\mu f  \partial_\nu  g + {1 \over 2!}  {\left( i \over 2 \right)^2} \theta^{\mu_1\nu_1} \theta^{\mu_2\nu_2}(\partial_{\mu_1}   \partial_{\mu_2} f )(\partial_{\nu_1}   \partial_{\nu_2} g )+ \cdots\label{starproduct}\nonumber\\
\end{eqnarray}
\end{small}
with a constant $\theta^{\mu\nu}$. Then the usual product is
repalced by the Moyal product in the classical lagrangian density.

In this work we study the representation of Galilei group in
noncommutative space. Particulary, we analyze the exotic structure
of Galilei algebra, that appears in special case of $2+1$ dimensions
\cite{lev1,lev2}. This exotic structure shows us some physical
properties which are not shared by another dimensions, for instance,
it allows a natural association between noncommutative coordinates
and the second central extension of the algebra. In this sense, the
exotic extension of Galilei algebra is responsible  for the
emergence of noncommutative coordinates \cite{jack1, jack2}.

The paper is organized as follows. In Section 2 we construct the
noncommutative Galilei algebra in $2+1$ dimensions and write the
Sch\"odinger equation in noncommutative plane. As a result, we
derive the Ehrenfest Theorem in noncommutative coordinates. In
section 3 we study the linear potential in noncommutative plane.
Then in section 3 we work with the harmonic oscillator in the
noncommutative plane. In section 4 we present the anharmonic
oscillator using noncommutative coordinates in two dimensions. We
obtain a correction in the energy using perturbation theory. Finally
in the last section we present our concluding remarks.

\section{Noncommutative Galilei-Lie Algebra}

In this section, we study the noncommutative Galilei algebra in $2+1$ dimension. For this proposal,  we construct the operators
\begin{equation}\label{o1}
\widehat{x}_i=x_i-\frac{i}{2}\theta_{ij}\frac{\partial}{\partial x_j},
\end{equation}
\begin{equation}\label{o2}
\widehat{p}_{i}=i\hbar \frac{\partial}{\partial x_i},
\end{equation}
\begin{equation}\label{o3}
\widehat{K}_i=m\widehat{x}_i-t\widehat{p}_i,
\end{equation}
\begin{equation}\label{o4}
\widehat{L}=\widehat{x}_i\widehat{p}_{j}-\widehat{x}_j\widehat{p}_{i},
\end{equation}
and
\begin{equation}\label{o5}
\widehat{H}=\frac{1}{2m}\sum_{i=1}^{2}\widehat{p}_{k}^{2}.
\end{equation}
From this set of unitary operators we obtain, after some long but simple calculations, the following set of commutation relations,
\begin{equation}\nonumber
[\widehat{p}_{i},\widehat{p}_{j}]=0,
\end{equation}
\begin{equation}\nonumber
[\widehat{K}_{i},\widehat{p}_{j}]=i\hbar m\delta_{ij},
\end{equation}

\begin{equation}\nonumber
[\widehat{L},\widehat{p}_{i}]=i\hbar\epsilon_{ij}\widehat{p}_{j},
\end{equation}
\begin{equation}\nonumber
[\widehat{K}_{i},\widehat{K}_{j}]=im^2\theta_{ij},
\end{equation}
\begin{equation}\nonumber
[\widehat{p}_{i},\widehat{H}]=0,
\end{equation}
\begin{equation}\nonumber
[\widehat{L},\widehat{H}]=0.
\end{equation}
This is the extended exotic Galilei algebra, where $\widehat{L}_{i}$ stand for rotations, $\widehat{p}_{i}$ for translations and $\widehat{K}_{i}$ for boosts. The central extension is given by $[\widehat{K}_{i},\widehat{p}_{j}]=i\hbar m\delta_{ij}$. However, in the plane, the Galilei group admits a second extension, highlighted by the non-commutativity of boost generators. This extension was studied first by Levy-Leblond \cite{lev1}, and this algebra is called the Exotic Galilei Group. In general, the second extension of Galilei algebra is associated to noncommutative coordinates. Here, we show that noncommutativity between coordinates leads to exotic extension.   To obtain the physical content, we first notice that $\widehat{p}$ transform under the boost as the physical momentum, i.e.,
\begin{equation}\nonumber
\exp\left(-i\overrightarrow{v}\cdot\frac{\widehat{K}}{\hbar}\right)\widehat{p}_{j}\exp\left(i\overrightarrow{v}\cdot\frac{\widehat{K}}{\hbar}\right)=\widehat{p}_j+mv_j\mathbf{1}.
\end{equation}
Futhermore, the operators $\widehat{x}$ and $\widehat{p}$ do not commute with each other, that is,
$$[\widehat{x}_i,\widehat{p}_j]=i\hbar\delta_{ij}.$$
Therefore, $\widehat{x}$ and $\widehat{p}$ can be taken to be the physical
observables of position and momentum. To be consistent,
generators $\widehat{L}$ are interpreted as the angular momentum operator,
and $\widehat{H}$ is taken as the Hamiltonian operator. The Casimir
invariants of the Lie algebra are given by
\begin{equation*}
I_{1}=\widehat{H}-\frac{\widehat{P}^{2}}{2m}\quad \mathrm{and}\quad I_{2}=%
\widehat{L}-\frac{1}{m}\widehat{K}\times \widehat{P},
\end{equation*}%
where $I_{1}$ describes the Hamiltonian of a free particle and $I_{2}$ is
associated with the spin degrees of freedom. First, we study the scalar
representation; i.e. spin zero.

Particularly, $\widehat{H}$ can be identify as generator of temporal translation. In this way, the time evolution of the system state $\psi(x,t)=\langle x|\psi(t)\rangle$ is given by
\begin{equation}\label{tt}
\psi(x,t)=e^{\frac{-i\widehat{H}t}{\hbar}}\psi(x,0).
\end{equation}
Differentiating Eq.(\ref{tt}) with respect to $t$, we obtain
\begin{equation}\label{es1}
i\hbar\partial_{t}\psi(x,t)=\frac{\widehat{p}^2}{2m}\psi(x,t)+V(x_i-\frac{i}{2}\theta_{ij}\partial_{x_j})\psi(x,t),
\end{equation}
which is the noncommutative Schr\"odinger equation. Note that Eq.(\ref{es1}) was obtained here from representation theory of  symmetry group.

As a important result, we derive the Ehrenfest theorem for the noncommutative coordinates. For this proposal, we take the expansion
$$V\left(x_k-\frac{1}{2\hbar}\theta_{kl}\widehat{p}_l\right)=V(x_k)-\frac{1}{2\hbar}\theta_{kl}\widehat{p}_l\frac{\partial V}{\partial x_k}+o(\theta^2),$$
and to use in the equation $\frac{d}{dt}\langle A\rangle=\frac{i}{\hbar}\langle [H,A]\rangle+\langle \frac{\partial A}{\partial t}\rangle$, we obtain
\begin{equation}\label{te1}
\frac{d}{dt}\langle \widehat{x}_k\rangle=\frac{\langle \widehat{p}_k\rangle}{m}-\frac{\theta_{kl}}{2\hbar}\langle \frac{\partial V}{\partial x_l}\rangle,
\end{equation}
\begin{equation}\label{te2}
\frac{d}{dt}\langle \widehat{p}_k\rangle=\langle -\frac{\partial V}{\partial x_k}\rangle+\frac{\theta_{jl}}{2\hbar}\langle \widehat{p}_l\frac{\partial^2 V}{\partial x_l\partial x_j}\rangle.
\end{equation}
Observe that if we take $\theta=0$, we obtain the same result for commutating coordinates. In sequence, we prove an important theorem valid for linear potential in noncommutative coordinates.
\section{Linear Potential}
In this section we analyze the linear potential in noncommutative coordinates. The problem of linear potential is applied for example in quantum chromodynamics for studies of gluon condensates (H. G. Dosch). Our focus is in two dimensional case $V(x,y)=\alpha x + \beta y$, where $\alpha$ and $\beta$ are complex constants. In this sense, the noncommutative Schr\"odinger equation with linear interaction can be written by ($\hbar=m=1$)
\begin{equation}\label{es2}
\frac{\partial^2\psi(x,y)}{\partial x^2}+\frac{\partial^2\psi(x,y)}{\partial y^2}+\alpha x\psi(x,y)+\beta y \psi(x,y)-i\alpha\frac{\theta}{2}\frac{\partial\psi(x,y)}{\partial y}+i\beta\frac{\theta}{2}\frac{\partial\psi(x,y)}{\partial x}=E\psi(x,y).
\end{equation}
Observe that we utilize the anti-symmetric tensor $\theta$ as
$$\theta_{ij}=\left[\begin{array}{rr}
0&\theta\\-\theta&0
\end{array}\right].$$
To solve Eq. (\ref{es2}) firstly we note that its imaginary part
yields the equation $$\beta\frac{\partial \psi(x,y)}{\partial
x}-\alpha\frac{\partial \psi(x,y)}{\partial y}=0\,,$$ which implies
$\psi=\psi(\alpha x+\beta y)$, thus we change variables with the
definition $z=(\alpha x+\beta y)$, such that $\frac{\partial
\psi}{\partial x}=\alpha\frac{\partial \psi}{\partial z}$,
$\frac{\partial \psi}{\partial y}=\beta\frac{\partial \psi}{\partial
z}$, $\frac{\partial^2 \psi}{\partial x^2}=\alpha^2\frac{\partial^2
\psi}{\partial z^2}$, $\frac{\partial^2 \psi}{\partial
y^2}=\beta^2\frac{\partial^2 \psi}{\partial z^2}$. This leads to
\begin{equation}\label{es3}
(\alpha^2+\beta^2)\frac{\partial^2\psi(z)}{\partial z^2}+z\psi(z)-E\psi(z)=0.
\end{equation}
Observe that Eq.(\ref{es3}) is of the general form
$$\frac{d^2u}{dx^2}-xu=0,$$
that is the Airy´s differential equation. So, the solutions of Eq.(\ref{es3}) are given by
\begin{equation}\label{linear}
\psi(x,y)=\frac{2^{2/3}}{2\pi}Ai(\alpha x+\beta y-E),
\end{equation}
where $Ai(z)$ é the Airy function. Observe that the solution does
not depend on the noncommutative parameter $\theta$, which shows
that the Linear Potential does not have noncommutative corrections.

\section{Two Dimensional Noncommutative Isotropic Harmonic Oscillator}

In this section we will deal with the two dimensional isotropic
harmonic oscillator in noncommutative coordinates. The Hamiltonian
for just for the 2D harmonic oscillator is given by

\begin{equation}\label{ho1}
\widehat{H}=\frac{1}{2m}\widehat{p}_x +\frac{1}{2m}\widehat{p}_y+\frac{1}{2}m(\omega_x^2 \widehat{x}^2+\omega_y^{2}\widehat{y}^2)\,,
\end{equation}
If we take $m=\hbar=1$ then the Hamiltonian reads

\begin{equation}\label{ho2}
\widehat{H}=\frac{1}{2m}\widehat{p}_x +\frac{1}{2m}\widehat{p}_y+\frac{1}{2}\omega_x^2 (x-\frac{i}{2}\theta_{12}\partial_{y})^2+\frac{1}{2}\omega_y^{2}(y-\frac{i}{2}\theta_{21}\partial_x)^2\,,
\end{equation}
Once we rearrange the terms it yields

\begin{align}
\widehat{H}&=-\frac{1}{2}\left[\left(1+\frac{\theta_{12}^{2}\omega_{y}^{2}}{4}\right)\partial^{2}_x+
\left(1+\frac{\theta_{12}^{2}\omega_{x}^{2}}{4}\right)\partial^{2}_y\right]+\nonumber\\
+& {}\frac{1}{2}(\omega_x^{2}x^2+\omega_y^{2}y^2)+\frac{i}{2}\theta_{12}(\omega_y^2 y\partial_x-\omega_x^2 x\partial_y)\,.\label{ho3}
\end{align}
Thus we would like to analyze the isotropic case which is settled by
the condition $\omega_x=\omega_y=\omega$. Performing a change of
variables $z=\frac{1}{2}(\omega_x^{2}x^2+\omega_y^{2}y^2)$, we
obtain

\begin{equation}\label{ho4}
\widehat{H}\psi=-\frac{1}{2}\left(1+\frac{\theta^2\omega^2}{4}\right)\left(2\omega^2\frac{d\psi}{dz}+2z\omega^2\frac{d^2\psi}{dz^2}\right)+z\psi\,,
\end{equation}
we recall that $\theta_{12}=-\theta_{21}=\theta$. Therefore the Schröedinger equation reads

\begin{equation}\label{ho5}
-\beta\left(z\frac{d^{2}\psi}{dz^2}+\frac{d\psi}{dz}\right)+z\psi=E\psi\,,
\end{equation}
where $\beta=\omega^2\left(1+\frac{\theta^2\omega^2}{4}\right)$. If we seek by solutions of the form
$\psi=e^{-\xi/\omega}\phi$, then we find

\begin{equation}\label{ho6}
\xi\frac{d^2\phi}{d\xi^2}+\left(1-\frac{2\xi}{\omega}\right)\frac{d\phi}{d\xi}+\frac{1}{\omega}\left(-1+\frac{\epsilon}{\omega}\right)\phi=0\,,
\end{equation}
where $\epsilon=E\omega/\sqrt{\beta}$. Performing another change of variables $\zeta=2\frac{\xi}{\omega}$ in order to obtain a non-dimensional equation, we get

\begin{equation}\label{ho7}
\zeta\frac{d^{2}\phi}{d\zeta^2}+(1-\zeta)\frac{d\phi}{d\zeta}+\frac{1}{2}\left(-1+\frac{\epsilon}{\omega}\right)\phi=0\,,
\end{equation}
the above equation is known as Kummer equation and its solution is the confluent hypergeometric function. If we look for polynomial solutions, then we get

\begin{equation}\label{ho8}
\psi(z)=c\exp(-z/\beta^{1/2})F(-n;1;2z/\beta^{1/2}),
\end{equation}
where $c$ is a normalization constant, n is an integer and $$\epsilon=\omega(2n+1).$$
Using the properties of the confluent hypergeometric functions, it is possible to write the solution as

\begin{small}
\begin{equation}\label{ho9}
\psi(x,y)=\left(\frac{\omega}{\beta^{1/4}2^{(n_1+n_2)/2}\sqrt{\pi
n_1!n_2!}}\right)\exp{\left[-\frac{\omega^2(x^2+y^2)}{2\beta^{1/2}}\right]}H_{n_1}\left(\frac{\omega
x}{\beta^{1/4}}\right)H_{n_2}\left(\frac{\omega
y}{\beta^{1/4}}\right)\,,
\end{equation}
\end{small}
where $n=\frac{(n_1+n_2)}{2}$ and $H_n(x)$ are the Hermite
polynomials. Thus the energy is given by $$E_{n_1\,n_2}=\left(1+\frac{\theta^2\omega^2}{4}\right)^{1/2}\omega\left(n_1+n_2+1\right)\,,$$
we can see that the correction due to the non-commutativity in the coordinates is given by the first factor of the above expression.

\section{Two Dimensional Noncommutative Anharmonic Oscillator}

In this section we will analyze the anharmonic oscillator in two
dimensions under, taking into account the feature of
non-commutativity between coordinates. This system consists of small
deviations from the normal oscillator. The Hamiltonian that
describes it just reads

\begin{equation}\label{hao1}
\widehat{H}=\frac{1}{2m}\widehat{p}_x
+\frac{1}{2m}\widehat{p}_y+\frac{1}{2}m(\omega_x^2
\widehat{x}^2+\omega_y^{2}\widehat{y}^2)+\alpha\left(\widehat{x}^3+\widehat{y}^3\right)+\gamma\left(\widehat{x}^4+\widehat{y}^4\right)\,,
\end{equation}
the quantities $\alpha$ and $\gamma$ are coupling constants, they
are supposed to be very small, in such a way that it is possible to
apply perturbation theory to calculate the corrections in the
energy. Thus given a Hamiltonian of the form
$\widehat{H}=\widehat{H}_0+\Delta\widehat{H}$ where
$\widehat{H}_0\Psi_0=E_0\Psi_0$, it is possible to evaluate the
correction by the expression $\Delta E=\left<\Psi_0\mid\Delta
\widehat{H}\mid\Psi_0\right>$. If we take $m=\hbar=1$, then we get
$$\widehat{H}_0=\frac{1}{2}\widehat{p}_x
+\frac{1}{2}\widehat{p}_y+\frac{1}{2}\omega_x^2
(x-\frac{i}{2}\theta_{12}\partial_{y})^2+\frac{1}{2}\omega_y^{2}(y-\frac{i}{2}\theta_{21}\partial_x)^2$$
and
\begin{small}
\begin{equation*}
\Delta\widehat{H}=\alpha\left[(x-\frac{i}{2}\theta_{12}\partial_{y})^3+(y-\frac{i}{2}\theta_{21}\partial_x)^3\right]
+\gamma\left[(x-\frac{i}{2}\theta_{12}\partial_{y})^4+(y-\frac{i}{2}\theta_{21}\partial_x)^4\right]\,.
\end{equation*}
\end{small}
From last section we identify $E_0$ as
$E_{n_1,n_2}=\sqrt{\beta}\left(n_1+n_2+1\right)$ and $\Psi_0$ as the
function into expression (\ref{ho9}).

In order to calculate the mean values of the quantities appearing in
the Hamiltonian, it is necessary to present some integrals. Thus
using the following relations
\begin{small}
\begin{eqnarray}
\int_{-\infty}^{\infty}
e^{-x^2}H^2_n(x)dx&=&2^nn!\sqrt{\pi}\,,\nonumber\\
\int_{-\infty}^{\infty}
xe^{-x^2}H_n(x)H_m(x)dx&=&\sqrt{\pi}2^nn!\left[\frac{1}{2}\delta_{n-1,m}+\left(n+1\right)\delta_{n+1,m}\right]\,,\nonumber\\
\int_{-\infty}^{\infty}
x^2e^{-x^2}H_n(x)H_m(x)dx&=&\sqrt{\pi}2^nn!\left[\left(n+\frac{1}{2}\right)\delta_{n,m}+\left(n+2\right)
\left(n+1\right)\delta_{n+2,m}+\frac{1}{4}\delta_{n-2,m}\right]\,,\nonumber
\end{eqnarray}
\end{small}
we get
\begin{eqnarray}
\int_{-\infty}^{\infty} x^3e^{-x^2}H^2_n(x)dx&=&0\,,\nonumber\\
\int_{-\infty}^{\infty}
x^3e^{-x^2}H_n(x)H_{n-1}(x)dx&=&3\sqrt{\pi}\,2^{n-2}n^2(n-1)!\,,\nonumber\\
\int_{-\infty}^{\infty}
x^4e^{-x^2}H^2_n(x)dx&=&3\sqrt{\pi}\,2^{n-2}(2n^2+2n+1)n!\,,\nonumber\\
\int_{-\infty}^{\infty}e^{-x^2/2}H_n(x)\left(\frac{\partial^2}{\partial
x^2}\right)e^{-x^2/2}H_n(x)
dx&=&\sqrt{\pi}2^n\left(n-\frac{1}{2}\right)n!\,,\nonumber\\
\int_{-\infty}^{\infty}e^{-x^2/2}H_n(x)\left(\frac{\partial^4}{\partial
x^4}\right)e^{-x^2/2}H_n(x)
dx&=&\frac{3}{2}\sqrt{\pi}2^n\left(3n^2-7n+\frac{1}{2}\right)n!\,.\nonumber
\end{eqnarray}
Hence we have
\begin{eqnarray}
\left<x^4\right>&=&\frac{3\beta}{2\omega^{7/2}}\left(n_1^2+n_1+\frac{1}{2}\right)\,,\nonumber\\
\left<y^4\right>&=&\frac{3\beta}{2\omega^{7/2}}\left(n_2^2+n_2+\frac{1}{2}\right)\,,\nonumber\\
\left<x^2\,\frac{\partial^2}{\partial
y^2}\right>&=&\left(n_1+\frac{1}{2}\right)\left(n_2-\frac{1}{2}\right)\,,\nonumber\\
\left<y^2\,\frac{\partial^2}{\partial
x^2}\right>&=&\left(n_1-\frac{1}{2}\right)\left(n_2+\frac{1}{2}\right)\,,\nonumber\\
\left<\frac{\partial^4}{\partial
y^4}\right>&=&\frac{3\omega^4}{2\beta}\left(3n_2^2-7n_2+\frac{1}{2}\right)\,,\nonumber\\
\left<\frac{\partial^4}{\partial
x^4}\right>&=&\frac{3\omega^4}{2\beta}\left(3n_1^2-7n_1+\frac{1}{2}\right)\,.\nonumber
\end{eqnarray}
Therefore the above expressions lead to the following results
\begin{small}
\begin{eqnarray}
\left<\widehat{x}^3\right>&=&0\,,\nonumber\\
\left<\widehat{y}^3\right>&=&0\,,\nonumber\\
\left<\widehat{x}^4\right>&=&\frac{3\beta}{2\omega^{7/2}}\left(n_1^2+n_1+\frac{1}{2}\right)-
\frac{3\theta^2}{2}\left(n_1+\frac{1}{2}\right)\left(n_2-\frac{1}{2}\right)+
\frac{3\theta^4\omega^4}{32\beta}\left(3n_2^2-7n_2+\frac{1}{2}\right)\,,\nonumber\\
\left<\widehat{y}^4\right>&=&\frac{3\beta}{2\omega^{7/2}}\left(n_2^2+n_2+\frac{1}{2}\right)-
\frac{3\theta^2}{2}\left(n_1-\frac{1}{2}\right)\left(n_2+\frac{1}{2}\right)+
\frac{3\theta^4\omega^4}{32\beta}\left(3n_1^2-7n_1+\frac{1}{2}\right)\,.\nonumber
\end{eqnarray}
\end{small}
Then the first order correction of the energy is given by
\begin{align}
\left<\Delta E\right> &= \gamma \left\{   \frac{3\beta}{2\omega^{7/2}}\left(n_1^2+n_2^2+n_1+n_2+1\right)- \frac{3\theta^2}{2}\left[\left(n_1+\frac{1}{2}\right)\left(n_2-\frac{1}{2}\right) +\right.\right.\nonumber\\
   &+\left.\left.{}\left(n_1-\frac{1}{2}\right)\left(n_2+\frac{1}{2}\right)\right]+\frac{3\theta^4\omega^4}{32\beta}\left[3\left(n_2^2+n_1^2\right)-7\left(n_2+n_1\right)+1\right]\right\}\,.
\end{align}
We stress out that the corrected energy has, itself, correction in
terms of the noncommutative parameter $\theta$. This means that we
recover what is known for the anharmonic oscillator plus a new
behavior due to the non-commutativity between coordinates.

\section{Concluding Remarks}

In this paper we have dealt with the noncommutative Galilei group
which is also known as Exotic Galilei group. Such a group only has
consistency when we work with two spatial dimensions and one
temporal dimension (2+1). In this framework we obtain the
noncommutative Schr\"odinger equation from a group structure rather
than the well known procedure consisting on to change the normal
product by the Moyal product into the Lagrangian. Then as
applications we analyze the noncommutative Schr\"odinger equation or
the case of a linear potential, harmonic oscillator and anharmonic
oscillator. We find no corrections for the first case. On the other
hand we give corrections for the energy in terms of the
noncommutative parameter $\theta$ for both oscillators (harmonic and
anharmonic). We point out that all corrections obtained in this
paper goes to the known quantities in the limit
$\theta\rightarrow\infty$.

\end{document}